\documentclass[aps,prl,reprint,amsmath,amssymb,superscriptaddress]{revtex4-2}
\usepackage[bookmarks=false,linkcolor=blue,urlcolor=blue,colorlinks,citecolor=blue]{hyperref}
\usepackage{graphicx}
\newcommand{\figref}[2]{\hyperref[#1]{\ref{#1}(#2)}}

\begin{document}

\title{Three-phase Majorana zero modes at tiny magnetic fields}

\author{Omri Lesser}
\email{omri.lesser@weizmann.ac.il}
\affiliation{Department of Condensed Matter Physics, Weizmann Institute of Science, Rehovot, Israel 7610001}

\author{Karsten Flensberg}
\affiliation{Center for Quantum Devices, Niels Bohr Institute, University of Copenhagen, DK-2100 Copenhagen, Denmark}

\author{Felix von Oppen}
\affiliation{Dahlem Center for Complex Quantum Systems and Fachbereich Physik, Freie Universit{\"a}t Berlin, Arnimallee 14, 14195 Berlin, Germany}

\author{Yuval Oreg}
\affiliation{Department of Condensed Matter Physics,    Weizmann Institute of Science, Rehovot, Israel 7610001}

\begin{abstract}
Proposals for realizing Majorana fermions in condensed matter systems typically rely on magnetic fields, which degrade the proximitizing superconductor and plague the Majoranas' detection.
We propose an alternative scheme to realize Majoranas based only on phase-biased superconductors. 
The  phases (at least three of them) can be biased by a tiny magnetic field threading macroscopic superconducting loops, focusing and enhancing the effect of the magnetic field onto the junction, or by supercurrents.
We show how a combination of the superconducting phase winding and the spin-orbit phase induced in closed loops (Aharonov-Casher effect) facilitates a topological superconducting state with Majorana end states. We demontrate this scheme 
by an analytically tractable model as well as simulations of realistic setups comprising only conventional materials.
\end{abstract}
\maketitle

\emph{Introduction.---}The realization of robust Majorana zero modes (MZMs) at the ends of quasi-one-dimensional (1D) $p$-wave superconductors (SCs) has been a long-standing goal in contemporary condensed matter physics~\cite{alicea_new_2012}.
These exotic quasiparticles, predicted to possess non-Abelian exchange statistics, signal the appearance of a novel phase of matter: a topological superconductor.
Interest in realizing MZMs has been stimulated by the fundamental-physics quest of discovering new phases of matter, as well as by potential applications to topological quantum computation~\cite{nayak_non-abelian_2008,OregReview2020}.

Following the canonical toy model of a spinless $p$-wave SC chain~\cite{kitaev_unpaired_2001}, several proposals for experimentally realizing MZMs have been put forward~\cite{lutchyn_majorana_2018}.
These platforms include the surface of topological insulators proximity coupled to a superconductor~\cite{fu_superconducting_2008}, hybrid semiconductor-superconductor nanowires~\cite{lutchyn_majorana_2010,oreg_helical_2010}, possibly current-biased~\cite{romito_manipulating_2012}, semiconductor-ferromagnet heterostructures~\cite{sau_generic_2010,vaitiekenas_zero-bias_2020}, quantum wells with an in-plane magnetic field~\cite{alicea_majorana_2010}, phase-biased Josephson junctions~\cite{potter_anomalous_2013,hell_two-dimensional_2017,pientka_topological_2017,ren_topological_2019,fornieri_evidence_2019}, carbon nanotubes~\cite{sau_topological_2013,marganska_majorana_2018,lesser_topological_2020}, chains of magnetic adatoms on superconductors with strong spin-orbit coupling~\cite{nadj-perge2014,pientka_topological_2013,nadj-perge_observation_2014}, and full-shell proximitized nanowires~\cite{stanescu_robust_2018,vaitiekenas_flux-induced_2020}.

Generally, three ingredients are needed to realize topological superconductivity in one dimension: proximity coupling to a conventional $s$-wave superconductor (sufficiently thick to be free of phase fluctuations), a spin-rotation mechanism, most commonly spin-orbit coupling (SOC), and a source of time-reversal-symmetry breaking.
With a proper combination of these ingredients, the low-energy band becomes effectively spinless while remaining susceptible to pairing, thus realizing a $p$-wave superconductor.
Time-reversal symmetry is usually broken by an external Zeeman field or by internal magnetic phenomena, such as the exchange field of a nearby ferromagnet.

Realizations in which the proximitizing superconductor is subjected to a magnetic field have the drawback of degrading superconductivity~\cite{sabonis_destructive_2020}.
In particular, all types of time-reversal-symmetry breakers -- Zeeman field, exchange field, magnetic flux in the presence of conventional impurities, or magnetic impurities -- lead to depairing of Cooper pairs and the formation of in-gap states.
In extreme cases, a gapless superconductor is formed~\cite{tinkham_introduction_2004}.
This makes MZMs fragile and renders their detection ambiguous.
Moreover, one may wonder why a Zeeman or exchange field is necessary at all.
Indeed, several proposals rely on controlling the phase of the SC order parameter only~\cite{fu_superconducting_2008,melo_supercurrent-induced_2019}. Other proposals include on top of that the application of a weak magnetic field~\cite{romito_manipulating_2012,hell_two-dimensional_2017,pientka_topological_2017,laeven_enhanced_2020,vaitiekenas_flux-induced_2020,kotetes_topological_2015}.

\begin{figure}
    \centering\includegraphics{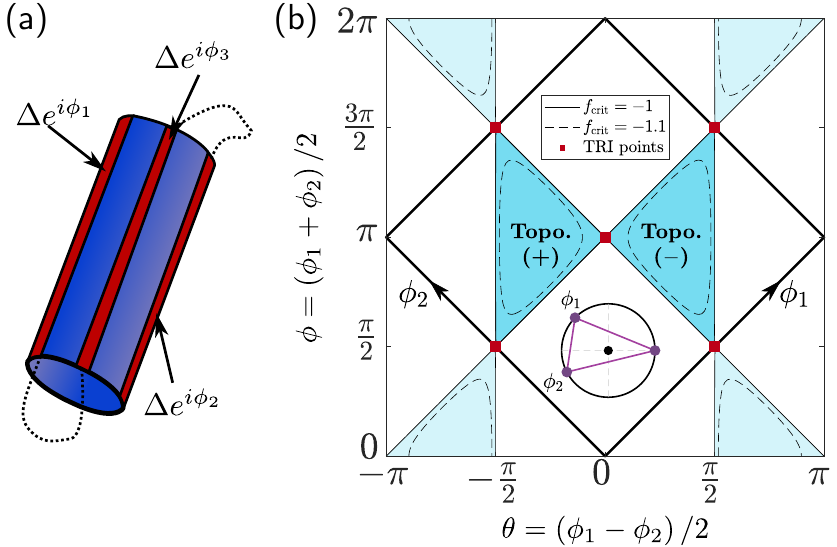}
    \caption{Coupled-wires model for topological superconductivity induced by phase bias only. 
    (a)~Illustration of the system under study: three spin-orbit-coupled wires, proximity coupled to SCs with three different phases. 
    (b)~Topological phase diagram as a function of the SC phases $\phi_1$ and $\phi_2$, with $\phi_3$ set to $0$ (see inset). The square in solid bold lines repeats periodically. 
    The solid thin lines correspond to the phase boundaries at the optimal manifold where the critical value required for a topological phase is $f_{\rm crit}=-1$, whereas the dashed lines correspond to $f_{\rm crit}=-1.1$, away from the optimal manifold. The phase windings in the central triangles realize a vortex ($+$) and an anti-vortex ($-$). Time-reversal-invariant points are shown as red squares. Notice that the phase diagram is  mirror symmetric about the line $\theta=0$.}
    \label{fig:cylinder}
\end{figure}

In this manuscript, we show that in the presence of a winding superconducting phase, topological superconductivity arises without any Zeeman field or magnetic flux penetrating the sample, using a conventional (non-topological) semiconducting substrate with strong spin-orbit coupling.
The distinction between opposite spins is generated by closed electron trajectories (loops) having gauge-invariant Aharonov-Casher phases~\cite{aharonov_topological_1984}. Such gauge-invariant phases arise when the loops encircle a net charge~\cite{aharonov_topological_1984}.
The winding can be obtained when the phases of at least three superconductors form a polygon on the unit circle surrounding the origin~\cite{van_heck_single_2014} [see Fig.~\figref{fig:cylinder}{b}].
This alleviates the need for a Zeeman field, an exchange field, magnetic fluxes~\cite{vaitiekenas_flux-induced_2020}, or relatively large supercurrents~\cite{laeven_enhanced_2020}.
The superconducting phases can be controlled by macroscopic superconducting loops, which focus the time-reversal-breaking element on the junction.
Therefore, a tiny magnetic field, of less than a micro-tesla for a micron-size loop, can be used to achieve topological superconductivity. In this method, the superconductors remain free of pair-breaking perturbations, in-gap states, and flux trapping, thereby allowing even the use of type-II superconductors such as Nb.

Fu and Kane~\cite{fu_superconducting_2008} studied the 2D surface of a 3D topological insulator in proximity to a conventional $s$-wave superconductor, showing that a discrete vortex associated with three phase-biased superconductors binds a MZM. This MZM emerges as a result of two topological phases accumulated by the surface Dirac fermions along loops encircling the vortex center. In addition to the $\pi$ phase associated with the vortex, there is a $\pi$ Aharonov-Casher (or Berry) phase~\cite{aharonov_topological_1984}, which originates from spin-momentum locking in the language of surface Dirac electrons. Importantly, both phases are required for inducing Majorana zero modes, even though the system is already topological without the vortex.
It would be very attractive to implement a similar scheme in 1D using conventional materials, as inducing a discrete vortex requires only minimal magnetic fields or supercurrents. However, it is not evident whether this is possible. First, the 1D system will not already be topological by proximity coupling to the conventional superconductor, and MZMs are in one-to-one correspondence with the formation of a topological superconducting state. Second, phase biasing does not directly introduce a Zeeman splitting, which is typically required. Here, we show for explicit examples that Aharonov-Casher phases in conjunction with a discrete vortex can stabilize topological superconductivity in 1D systems using conventional materials. We believe that this design principle can be highly beneficial in realizing topological superconductivity, as it eliminates the severely detrimental effects of large magnetic fields.

Our main result is the phase diagram in Fig.~\figref{fig:cylinder}{b} for the three-phase system depicted in Fig.~\figref{fig:cylinder}{a}.
The phase diagram depends on the two phase differences $\phi_1$ and $\phi_2$ ($\phi_3$ is set to zero), and periodically repeats the unit cell
indicated by the black square~\footnote{\label{note:timereversal}The phases are conveniently visualized by plotting them on a unit circle, fixing the global phase of the superconductors such that $\phi_3=0$. Then, the phases form a triangle as shown in the inset of Fig.~\figref{fig:cylinder}{b}.
When all $\phi_n=0$, the system is obviously time-reversal symmetric. At the three additional time-reversal-symmetric points $(\phi_1,\phi_2)=(0,\pi),(\pi,0),(\pi,\pi)$, the triangle degenerates into a line through the center of the unit circle.}.
To highlight the role of time-reversal symmetry,
we plot the phase diagram as a function of $\theta=\left(\phi_1-\phi_2\right)/2$ and $\phi=\left(\phi_1+\phi_2\right)/2$. 
Then, similar to a single phase-biased planar Josephson junction~\cite{hell_two-dimensional_2017,pientka_topological_2017}, $\phi=\pi$ and $\theta=0$ is a time-reversal-symmetric point  (as are $\phi=\pi/2,\theta=\pi/2$ and $\phi=3\pi/2,\theta=\pi/2$).
In contrast to conventional Josephson junctions where a Zeeman field is needed to break time-reversal symmetry and to drive the system into a topological state, here this effect is achieved by the phase difference $\theta$ between the superconductors.

\emph{Coupled-wires model.---}To demonstrate our approach in a tractable model, we consider three spin-orbit-coupled wires in proximity to three $s$-wave superconductors with pair potentials of magnitude $\Delta$ and phases $\phi_1,\phi_2,\phi_3$, as illustrated in Fig.~\figref{fig:cylinder}{a}.

In the continuum limit, the topological properties are already encoded in the spectrum for zero momentum along the wires, $k_{\parallel}=0$ \cite{kitaev_unpaired_2001}. 
In this case, the Hamiltonian of the three-wire model takes the form
\begin{equation}\label{eq:H_cylinder_kpara0_TB}
\begin{aligned}\mathcal{H}\left(k_{\parallel}=0\right)=& \sum_{n=1}^{N}\sum_{s,s'=\pm}\left[-\mu\delta^{ss'}c_{n,s}^{\dagger}c_{n,s'}\right.\\
&+ \left.\left(t_{\perp}\left(e^{i\lambda_{n}\sigma_{z}}\right)^{ss'}c_{n,s}^{\dagger}c_{n+1,s'}+{\rm H.c.}\right)\right]\\
& +\sum_{n=1}^{N}\left(\Delta e^{i\phi_{n}}c_{n,\uparrow}^{\dagger}c_{n,\downarrow}^{\dagger}+{\rm H.c.}\right),
\end{aligned}
\end{equation}
where $c_{n,s}$ annihilates an electron in wire $n$ with $k_{\parallel}=0$ and spin projection $s$ along $z$, $t_{\perp}$ is the inter-wire hopping amplitude, $\mu$ is the chemical potential, $\Delta$ is the induced SC pair potential, and $\lambda_n$ is the SOC angle accumulated between the neighboring wires $n$ and $n+1$. Here, we assume
periodic boundary conditions, $c_{N+1,s}=c_{1,s}$. 
As we will see, it is crucial that electrons acquire an Aharonov-Casher phase~\cite{aharonov_topological_1984}, which will conspire with the SC phase winding to eliminate one spin species at the Fermi level.
Equation \eqref{eq:H_cylinder_kpara0_TB} is written for a general number of wires $N$; for simplicity, we will focus on the minimal value to create a phase winding, $N=3$.
Notice that the gauge transformation 
$c_{n,s}\rightarrow c_{n,s} e^{i\phi_{n}/2}$ 
eliminates the phases from the SC terms and changes the hopping term to 
$t_{\perp}\rightarrow t_{\perp}\exp\left(i\frac{\phi_{n+1}-\phi_{n}}{2}\right)$.
This resembles but is not equivalent to magnetic flux: unlike magnetic flux, the phases $\phi_n$ can be gauged away when $\Delta=0$.

To identify phase transitions in the parameter space of our model, we search for gap closures by equating the determinant of the Hamiltonian Eq.~\eqref{eq:H_cylinder_kpara0_TB} to zero:
\begin{equation}\label{eq:det_H_kpara0}
\begin{aligned}
\det \mathcal{H}&\left(k_{\parallel}=0\right)=6\mu^{2}t_{\perp}^{2}\left(\Delta^{2}+\mu^{2}\right) -\left(\Delta^{2}+\mu^{2}\right)^{3} \\
&-3t_{\perp}^{4}\left(\Delta^{2}+3\mu^{2}\right) -2f\Delta^{2}t_{\perp}^{2}\left(\Delta^{2}+\mu^{2}+t_{\perp}^{2}\right) \\
&-4\mu t_{\perp}^{3}\Lambda\left(f\Delta^{2}-\mu^{2}+3t_{\perp}^{2}\right) -4t_{\perp}^{6}\Lambda^2 = 0,
\end{aligned}
\end{equation}
where $f=\cos\left(\phi_{1}-\phi_{2}\right)+\cos\left(\phi_{2}-\phi_{3}\right)+\cos\left(\phi_{3}-\phi_{1}\right)$ and $\Lambda=\cos\left(\lambda_1+\lambda_2+\lambda_3\right)$.

The SC phases appear in the determinant through a single parameter $-3/2\leq f \leq 3$, which has a simple geometric interpretation: for $-3/2\leq f \leq -1$ the phases wind, i.e., when plotted as complex numbers $\left\{e^{i\phi_n}\right\}$ on the unit circle, the triangle connecting them contains the origin~\cite{SupplementalMaterial}.
Solving the quadratic equation $\det\mathcal{H}\left(k_{\parallel}=0\right)=0$ for $\Lambda$, we find that a real solution is possible only for $f\leq -1$~\cite{SupplementalMaterial}, and therefore phase winding is a necessary condition for the existence of a zero-energy state, in agreement with the results of Ref.~\cite{van_heck_single_2014}.

Assuming that the SC phase winds, we still have to determine the regions in the three-dimensional parameter space spanned by $\mu$, $\Delta$, and $\Lambda$ (choosing units such that $t_\perp=1$) for which the system is topological.
An optimal situation occurs when the values of the three parameters are such that the determinant Eq.~\eqref{eq:det_H_kpara0} is zero already for $f=-1$.
Then, the system is topological for the maximal range of $-3/2\le f<-1$.
Setting $f=-1$ in Eq.~\eqref{eq:det_H_kpara0} we find that the optimal situation occurs when $\left(\mu,\Delta,\Lambda\right)$ are points on a circle ${\cal C}$ parametrized by $\left(\mu, \sqrt{1-\mu^2},\mu\right)$, see \cite{SupplementalMaterial} and Fig.~\ref{sfig:f_manifold}.
Setting $f=f_{\rm crit}$ with $-3/2 \le f_{\rm crit}<-1$ in Eq.~\eqref{eq:det_H_kpara0} defines a surface in the parameter space; when $\left( \mu, \Delta, \Lambda \right)$ lie on this surface, topological superconductivity occurs for $-3/2 \le f<f_{\rm crit}$, see Fig.~\figref{fig:cylinder}{b}.
Hence we conclude that topological superconductivity is obtained for all $\left(\mu, \Delta, \Lambda\right)$ points within the bulk of the shape defined at $f_{\rm crit}=-3/2$ (see Fig.~\ref{sfig:f_manifold} of the Supplemental Material~\cite{SupplementalMaterial}), with optimal values on the circle ${\cal C}$.

To find the energy gap in the topological state, we analyze the full spectrum of the system away from $k_{\parallel}=0$. Belonging to symmetry class D~\cite{altland_nonstandard_1997,schnyder_classification_2008,kitaev_periodic_2009}, the full Hamiltonian is characterized by the $\mathbb{Z}_2$ topological invariant~\cite{kitaev_unpaired_2001,lutchyn_search_2011}
\begin{equation}\label{eq:pfaffian}
    \mathcal{Q} = {\rm sign} \left[ {\rm Pf}\left( \mathcal{P} \mathcal{H}(k_{\parallel}=0) \right) {\rm Pf}\left( \mathcal{P} \mathcal{H}(k_{\parallel}=\pi) \right) \right],
\end{equation}
where Pf is the Pfaffian and $\mathcal{P}$ is the particle-hole operator.
$\mathcal{Q}=1$ indicates the trivial phase, whereas $\mathcal{Q}=-1$ in the topological phase, where the system supports MZMs~\footnote{In the continuum limit, the parallel part of the Hamiltonian reads $\mathcal{H}_{\parallel}=\left({k_{\parallel}^2}/{2m^*} + u k_{\parallel}\sigma_x\right)\tau_z$, where $m^*$ is the effective electron mass, $u$ is the Rashba SOC parameter along the wires, and the Pauli matrix $\tau_z$ acts in particle-hole space.
}.
The energy gap must be calculated for all values of $k_{\parallel}$.

\begin{figure}[ht]
    \centering\includegraphics{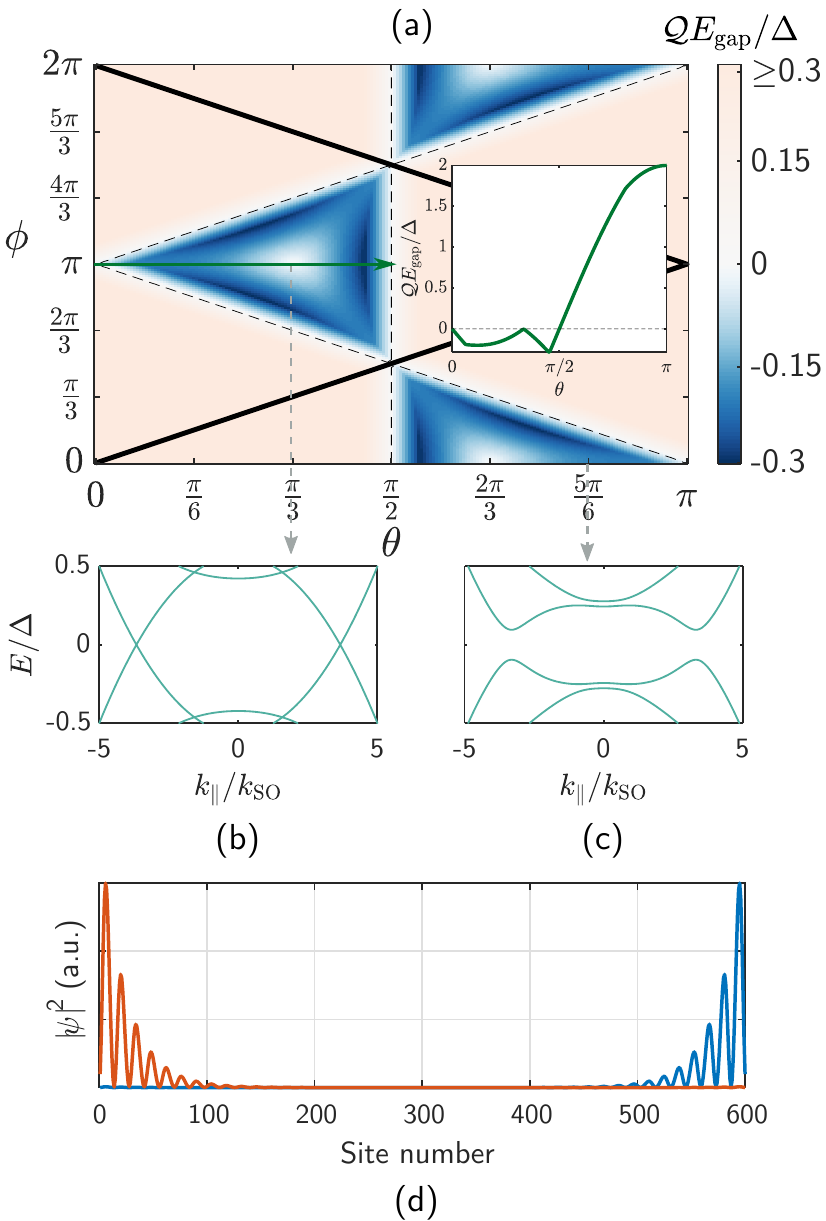}
    \caption{(a)~Topological phase diagram of the coupled-wires model as a function of the SC phase differences $\theta=\left(\phi_1-\phi_2\right)/2$ and $\phi=\left(\phi_1+\phi_2\right)/2$ (we set $\phi_3=0)$. The color scale shows the $\mathbb{Z}_2$ invariant $\mathcal{Q}$, which is $+1$ ($-1$) in the trivial (topological) phase, multiplied by the energy gap (normalized by $\Delta$) . The dark blue regions correspond to a robust large-gap topological phase. The phase boundaries (dashed lines) and Brillouin zone boundaries (solid lines) are marked. The inset shows a cut at $\phi=\pi$.
    The parameters are $t_{\perp}=1$, $\Delta=0.1$, $\mu=0.995$, $\lambda=0.033$ (on the optimal manifold), $m=0.01$, $u=1$.
    (b)~At the $C_3$-symmetric point $\theta=\frac{\pi}{3}$, $\phi=\pi$, the system becomes gapless at finite $k_{\parallel}$. 
    (c)~The gap closing becomes an avoided crossing when the $C_3$ symmetry is broken, done here by changing the phases away from the $C_3$-symmetric point.
    (d)~Wavefunctions of (near) zero-energy Majorana states in the topological phase, calculated for an open system discretized with $L=600$ sites per wire, at $\left(\theta=\frac{5\pi}{6},\phi=0\right)$.}
    \label{fig:finite_k_para}
\end{figure}

The numerically calculated~\cite{wimmer_algorithm_2012} topological phase diagram of the system is shown in Fig.~\figref{fig:finite_k_para}{a}, for parameters chosen on the optimal manifold.
Remarkably, the model supports a topological phase with an excitation gap of about $0.3\Delta$, with the application of only a phase difference and without any applied Zeeman or orbital field in the sample.
We note that at $\left(\theta=\frac{\pi}{3},\phi=\pi\right)$ and $\left(\theta=\frac{2\pi}{3},\phi=0\right)$ -- perfect vortices with equal phase differences forming an equilateral triangle -- the system becomes $C_3$-symmetric and turns out to be gapless, due to a non-topological gap closure at finite $k_{\parallel}$, see Fig.~\figref{fig:finite_k_para}{b}. In Fig.~\figref{fig:finite_k_para}{c}, we demonstrate that the gap opens when the $C_3$ symmetry is broken~\footnote{
This delicate gap closure originates from the high symmetry of the system: For $\lambda_1=\lambda_2=\lambda_3\equiv \lambda$, there is a $C_3$ rotation symmetry at these points which is removed by any symmetry-breaking perturbation, such as adding a non-proximitized wire, depleting one of the wires, or choosing different $\lambda_n$'s.}.
In addition, the topological gap is bounded from above by the minimum of $\Delta$ (the induced SC gap) and $\Delta_{\rm SO} \sim t_\perp \sin^2\left(\frac{1}{2N}\sum_n \lambda_n\right)$ (the SOC splitting energy).

Further confirmation for the existence of the topological phase is given in Fig.~\figref{fig:finite_k_para}{d}, where we show the Majorana wavefunctions deep in the topological phase.
These are obtained by diagonalizing the Hamiltonian on a finite lattice.
The appearance of two localized Majorana modes with near zero energy at the edges of the system signals its topological nature.

\begin{figure}
    \centering\includegraphics{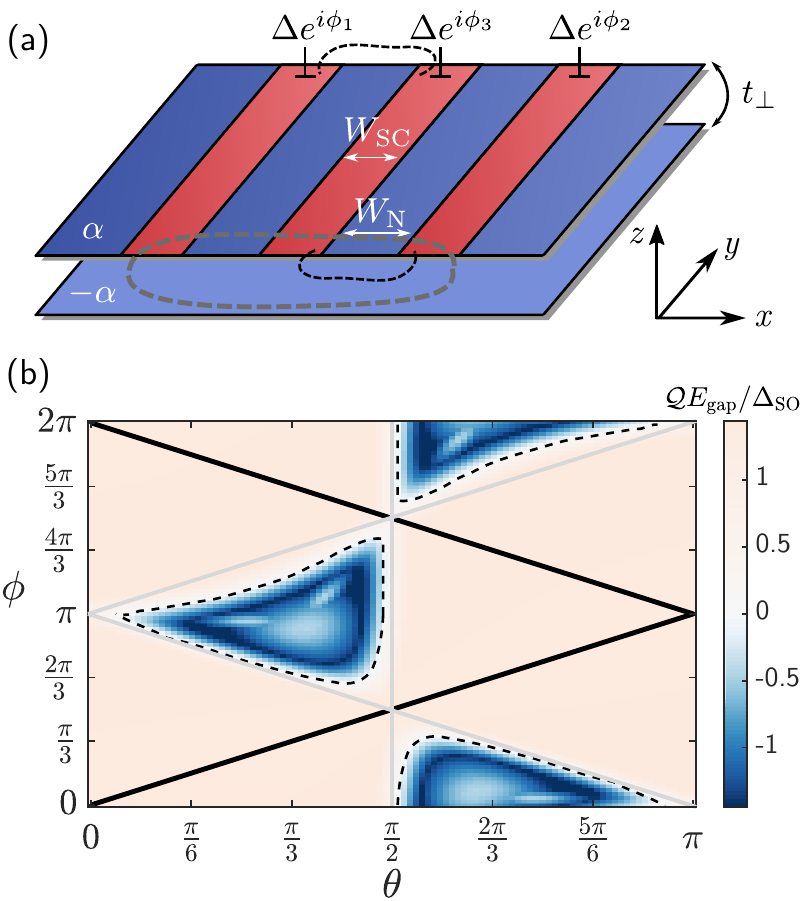}
    \caption{Quantum-well model for topological superconductivity induced by phase bias only. 
    (a)~Schematic of the experimentally available proposal: a spin-orbit-coupled two-layer 2DEG is contacted by three SCs of width $W_{\rm SC}$, separated by normal regions of width $W_{\rm N}$.
    The Rashba SOC parameter $\alpha$ is assumed to be opposite in the two layers, and pairing is only induced in one layer. The dashed gray line shows an example of a closed trajectory that encircles an Aharonov-Casher phase and is affected by the SC phase winding.
    (b)~Topological phase diagram of the InSb quantum-well model as a function of the SC phase differences $\theta=\left(\phi_1-\phi_2\right)/2$ and $\phi=\left(\phi_1+\phi_2\right)/2$ (setting $\phi_3=0)$.
    The color scale shows the product of the $\mathbb{Z}_2$ invariant $\mathcal{Q}$, which is $+1$ ($-1$) in the trivial (topological) phase, and the energy gap (normalized by the SOC energy $\Delta_{\rm SO}$). 
    Significant regions of $\mathcal{Q}=-1$ with a large energy gap appear (dark blue), implying a robust topological phase.
    The phase boundaries (dashed black lines), Brillouin zone boundaries (solid black lines), and optimal phase boundaries (gray lines) are marked.
    Parameters used: $\mu=108.9\,{\rm meV}$, $t_{\perp}=0.4\,{\rm meV}$, corresponding to a density of $n=6.4\times 10^{11}\,{\rm cm}^{-2}$ and a Fermi wavelength of $\lambda_{\rm F}=31\,{\rm nm}$.}
    \label{fig:quantum_well}
\end{figure}

\emph{Quantum-well model.---}Having established the possibility of realizing a 1D topological superconductor based on phase bias alone, we now turn to exemplifying this concept in a realistic system comprising readily available ingredients.
Specifically, our proposal relies on a spin-orbit-coupled 2DEG proximitized by three thick SCs.
As we have seen, the topological transition requires an Aharonov-Casher phase and thus, our proposal does not easily lend itself to an all-planar geometry.
Instead, we propose to use a 2DEG with two (or more) layers giving rise to several subbands, see Fig.~\figref{fig:quantum_well}{a}.
If the Rashba SOC parameter $\alpha$ is different in the two subbands, there are closed loops in which electrons acquire a non-zero Aharonov-Casher phase, mimicking the periodic boundary conditions in the simplified model we previously studied.

The system is described by the continuum Hamiltonian 
\begin{eqnarray}\label{eq:quantum_well}
\mathcal{H} & =&\left[-\frac{1}{2m^{*}}\left(\partial_{x}^{2}+\partial_{y}^{2}\right)-t_{\perp}\rho_{x}-\mu\right]\tau_{z}\\
 & +&i\alpha\left(\sigma_{x}\partial_{y}-\sigma_{y}\partial_{x}\right)\tau_z \rho_z +\left[\Delta\left(x\right)\tau_{+}+\Delta^{*}\left(x\right)\tau_{-}\right]\rho_{\uparrow}, \nonumber
\end{eqnarray}
where the Pauli matrices $\tau$, $\rho$ act in particle-hole and layer space, respectively, $t_{\perp}$ is the interlayer hopping amplitude, $\tau_{\pm}=\left(\tau_x\pm i\tau_y\right)/2$, and $\rho_{\uparrow}=(\rho_{0}+\rho_{z})/2$..
We assume that the Rashba SOC parameter is opposite in the two layers.
To be specific, we consider an InSb 2DEG with $m^*=0.014 m_{\rm e}$ and $\alpha=15\,{\rm meV\, nm}$~\cite{kallaher_spin-orbit_2010}, corresponding to a SOC length $\ell_{\rm SO}\approx 360\, {\rm nm}$.
We take an induced SC gap of $\Delta=1\,{\rm meV}$, appropriate for, e.g., Nb and Pb~\cite{matthias_superconductivity_1963}, in only one layer.
The widths of the SCs (normal regions between them) are chosen to be $W_{\rm SC}=70\,{\rm nm}$ ($W_{\rm N}=40\,{\rm nm}$.)
The typical length $W$ is chosen roughly according to the relation $\ell_{\rm SO}\Delta_{\rm SO}=W\Delta$.
This rule of thumb, which is derived in the Supplemental Material~\cite{SupplementalMaterial}, provides a way to approximate  favorable dimensions of the system given the material's parameters~\footnote{In our simulations we used $\Delta=1\,{\rm meV}$, a large value which makes a topological phase accessible in relatively small system. When using Al as the SC, the energy gap is about ten times smaller and the system would have to be about ten times larger.}.

The Hamiltonian Eq.~\eqref{eq:quantum_well} was investigated by discretizing it on a lattice of spacing $a=10\, {\rm nm}$.
The topological phase diagram, calculated by the Pfaffian formula Eq.~\eqref{eq:pfaffian} (now with $k_{\parallel}\,\widehat{=}\, k_y$), is shown in Fig.~\figref{fig:quantum_well}{b}.
The system indeed becomes a topological superconductor in the relevant region of phases.
The topological phase constitutes 17\% of the displayed $\theta$--$\phi$ section, compared to 25\% on the optimal manifold of the coupled-wires model [cf. Fig.~\figref{fig:cylinder}{b}], implying that further optimization is possible.
The maximal topological gap is of order $\Delta_{\rm SO}$, which is reasonable: for the chosen materials $\Delta_{\rm SO}$ is the smallest energy scale.
Using materials with larger $\Delta_{\rm SO}$ will lead to a larger topological gap.

As seen in Fig.~\figref{fig:quantum_well}{b}, the gap is small compared to $\Delta_{\rm SO}$ in some parts of the topological region.
By inspecting the Bogoliubov-de-Gennes spectrum, we find that the small gap originates from the presence of low-energy high-$k_{\parallel}$ modes.
Semiclassically, these modes result from long trajectories that hardly encounter the SCs, which is a common problem in such systems~\cite{laeven_enhanced_2020}.
Perturbations that eliminate these trajectories, such as non-standard geometries~\cite{laeven_enhanced_2020,melo_supercurrent-induced_2019} or disorder~\cite{haim_benefits_2019}, lead to an increased topological gap.
We have verified that adding a chemical potential modulation along the $x,y$ directions may significantly increase the topological gap. Furthermore, in the Supplemental Material~\cite{SupplementalMaterial} we show that the topological phase is robust to various perturbations in the model's parameters.

\emph{Discussion.---}In contrast to the vast majority of previous schemes, the topological phase in our proposal is induced solely by phase winding in the SC, which is proximity coupled to semiconductors with strong spin-orbit coupling such as InAs, InSb, or HgTe. 
SC phases can be manipulated using large external loops, through which magnetic flux is threaded, or by application of supercurrent.
The applied magnetic field (or the supercurrent), being very small and removed from the sample itself, should have only a mild effect on the parent SC.
Therefore in-gap states, which may mask the MZMs, are unlikely to appear.

We illustrated our scheme by an analytically accessible toy model and introduced a realistic setup in which these ideas can be implemented. Beyond these settings, we expect that the concept presented here -- relying  exclusively on SC phase bias and on the spin-dependent phase acquired in closed loops (the Aharonov-Casher phase~\cite{aharonov_topological_1984}) -- may be harnessed in other systems, as well.
For example, it might be possible to realize the wire model experimentally by contacting three of the six facets of an InAs nanowire with three thick phase-biased SCs.
The role of disorder deserves a separate treatment.
Disorder eliminates trajectories that do not encounter the superconductors~\cite{golubov_theoretical_1988,belzig_local_1996,pilgram_excitation_2000,haim_benefits_2019} and therefore increases the topological gap. We expect that under the right conditions it also gives rise to non-retro Andreev reflection, thereby facilitating the existence of the relevant closed trajectories.

Finally, a desirable goal for all Majorana platforms is an extension to networks to implement quantum information processing or a two-dimensional chiral phase~\cite{alicea_non-abelian_2011}. In our proposal, 
the experimental challenge is to establish control over a larger number of superconducting phases. At the same time, engineering aspects may be significantly simplified by the absence of a need for a Zeeman field, which requires careful alignment and induces harmful in-gap states.

\emph{Acknowledgment.---}We are grateful to C. M. Marcus, N. Schiller, G. Shavit, and A. Yacoby for fruitful discussions. K.F. acknowledges support from the Danish National Research Foundation.  F.v.O. is supported by Quantera-Grant
TOPOQUANT.

\bibliography{library}


\begin{center}
\large{\textbf{Supplemental Material}}
\end{center}

\setcounter{equation}{0}
\renewcommand{\theequation}{S\arabic{equation}}
\setcounter{figure}{0}
\renewcommand{\thefigure}{S\arabic{figure}}
\setcounter{section}{0}
\renewcommand{\thesection}{S\Roman{section}}

\section{Relation between $f$ and winding}
Here we prove the relation between $f\left(\phi_1,\phi_2,\phi_3\right)$, that appears in Eq.~\eqref{eq:det_H_kpara0} of the main text, and phase winding.
Without loss of generality, let us set $\phi_3=0$ and $\phi_2>\phi_1$, and examine 
\begin{equation}
    f\left(\phi_1,\phi_2,0\right) = \cos\left(\phi_1\right) + \cos\left(\phi_2\right) + \cos\left(\phi_2 - \phi_1\right).
\end{equation}
It is instructive to factor this expression using trigonometric identities:
\begin{widetext}
\begin{equation}\label{eq:f_trigo}
    \begin{aligned}
        f\left(\phi_1,\phi_2,0\right) &= 2\cos\left(\frac{\phi_1+\phi_2}{2}\right) \cos\left(\frac{\phi_2-\phi_1}{2}\right) + 2\cos^2\left(\frac{\phi_2-\phi_1}{2}\right)-1\\
        &= 2\cos\left(\frac{\phi_2-\phi_1}{2}\right) \left[ \cos\left(\frac{\phi_1+\phi_1}{2}\right) + \cos\left(\frac{\phi_2-\phi_1}{2}\right) \right] - 1 \\
        &= 4 \cos\left(\frac{\phi_2-\phi_1}{2}\right) \cos\left(\frac{\phi_1}{2}\right) \cos\left(\frac{\phi_2}{2}\right)-1.
    \end{aligned}
\end{equation}
\end{widetext}
The phases wind, i.e., the triangle connecting them encircles the origin, if and only if 
\begin{equation}\label{eq:winding_boundaries}
    0\leq \phi_1 \leq \pi,\quad \pi\leq \phi_2 \leq \pi+\phi_1.
\end{equation}
It follows that
\begin{equation}
    \cos\left(\frac{\phi_1}{2}\right)>0,\,
    \cos\left(\frac{\phi_2}{2}\right)<0,\,
    \cos\left(\frac{\phi_2-\phi_1}{2}\right)>0.
\end{equation}
Therefore, the first term in Eq.~\eqref{eq:f_trigo} is non-positive, and thus $f\leq-1$.
In addition, it is straightforward to show that along the boundaries defined by Eq.~\eqref{eq:winding_boundaries}, $f=-1$ exactly, and that $f\left(\phi_1,\phi_2,0\right)$ has extrema only at the points 
$\left(\phi_1,\phi_2\right)=\left(\frac{2\pi}{3},\frac{4\pi}{3}\right),\left(0,\pi\right),\left(\pi,\pi\right),\left(\pi,2\pi\right)$.
This concludes the proof that phase winding occurs if and only if $f\leq-1$.

\section{Further analysis of the phase diagram}
In this section we provide further details of the phase diagram in the coupled-wires model.
The derivations are based on Eq.~\eqref{eq:det_H_kpara0} of the main text, which determines the phase boundaries.

We begin by setting $t_{\perp}=1$, and writing the solution of Eq.~\eqref{eq:det_H_kpara0} which is a quadratic equation for $\Lambda$:
\begin{widetext}
\begin{equation}\label{eq:lsol}
\Lambda_{\pm}=\frac{1}{2}\left[\mu\left(\Delta^{2}f+3-\mu^{2}\right)\pm\left|\Delta\right|\sqrt{\mu^{2}\left[f\left(\Delta^{2}f+4\right)+6-3\Delta^{2}\right]-\left[\Delta^{2}(\Delta^{2}+2f)+(2f+3)\left(1+\mu^{4}\right)\right]}\right].
\end{equation}
\end{widetext}
In order to have a real solution, the argument of the square root must be non-negative. 
Treating the argument of the square root as a second-order polynomial in $\mu^2$, we find that the condition for a real solution is  
\begin{equation}
    (f-3) (f+1) \left(\Delta ^2 (f+1)^2+8 f+12\right) \geq 0.
\end{equation}
Since $-3/2\leq f \leq 3$, the first factor is negative whereas the last factor is positive. 
Therefore, this inequality is fulfilled only when $f\leq-1$.
This result, combined with our previous proof that $f\leq-1$ corresponds to phase winding, is in agreement with Ref.~\cite{van_heck_single_2014}.

\begin{figure*}
    \centering\includegraphics[width=\textwidth]{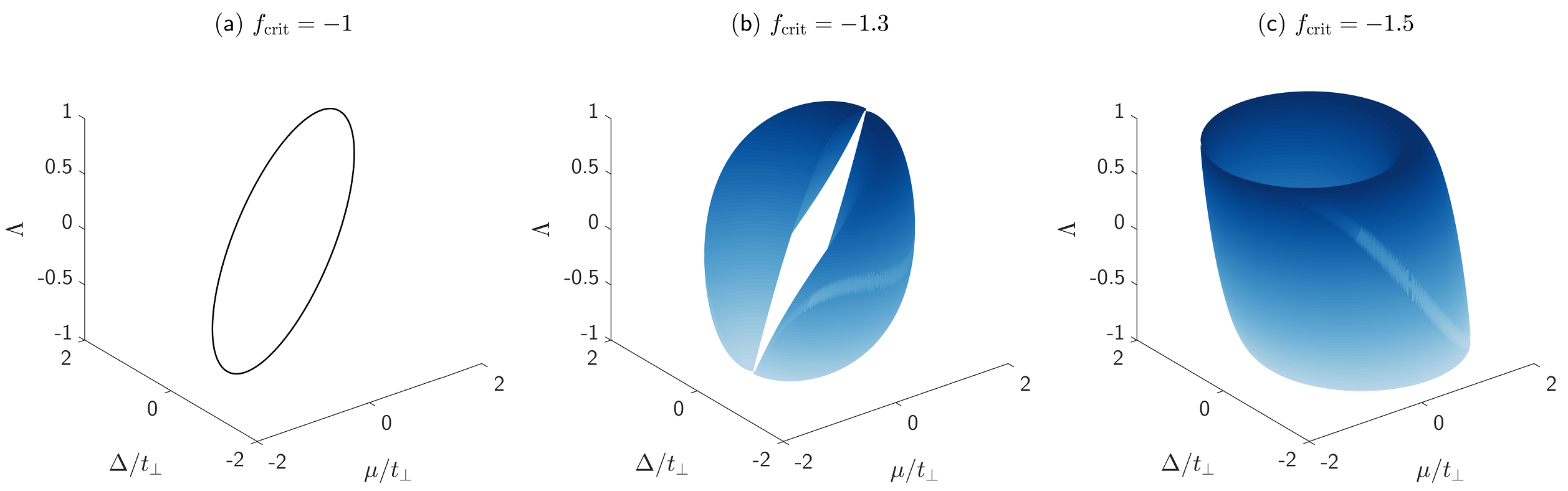}
    \caption{Critical manifold, marking the topological phase boundaries, in $\mu/t_{\perp}$, $\Delta/t_{\perp}$, $\Lambda$ space for (a)~$f_{\rm crit}=-1$, (b)~$f_{\rm crit}=-1.1$, (c)~$f_{\rm crit}=-1.5$, determined by the solution given in Eq.~\eqref{eq:lsol}. 
    As $f_{\rm crit}$ becomes more negative, the manifold's area increases. For a given value of $f_{\rm crit}$, the system is topological at all points contained in the volume surrounded by the surface.
    For $f_{\rm crit}=-1$ the manifold shrinks into a circle, as seen from Eq.~\eqref{eq:mu_lambda_cirlce}.
    Alternatively, fixing a point in the parameter space, we can contain it within the volume surrounded by the surface by changing $f$, leading to a topological state.}
    \label{sfig:f_manifold}
\end{figure*}

The manifold in $\mu$, $\Delta$, $\Lambda$ parameter space defined by Eq.~\eqref{eq:lsol} is shown in Fig.~\ref{sfig:f_manifold} for several values of $f$.
As explained in the main text, the parameters $\mu,\Delta,\Lambda$ are said to be ``optimal" if they support a solution of Eq.~\eqref{eq:det_H_kpara0} for $f=-1$, i.e., if the necessary condition $f\leq-1$ is also sufficient.
Solving again for $\Lambda$, we obtain 
\begin{equation}\label{eq:mu_lambda_cirlce}
    \Lambda_{\pm} = \frac{1}{2}\left[\mu  \left(3-\Delta ^2-\mu ^2\right) \pm \sqrt{-\Delta^2  \left(1-\Delta ^2-\mu ^2\right)^2}\right].
\end{equation}
The only way to make this expression real is demanding $\mu^2 + \Delta^2 = 1$.
In this case the two solutions $\Lambda_{\pm}$ are identical and equal to $\mu/t_{\perp}$.
This gives us the optimal curve $\mathcal{C}$ -- the circle $\left(\mu,\sqrt{1-\mu^2},\mu\right)$.

It is worth noting that along this circle, one can make a simple connection to the continuum description, thus finding an optimal condition for topological superconductivity in experimental system parameters. 
If we take $\Delta/t_{\perp}$ to be small (which means $\mu\approx t_{\perp}$), we obtain $\lambda\approx\Delta/t_{\perp}$ (assuming $\lambda_1=\lambda_2=\lambda_3=\lambda$). 
In our minimal three-wires tight-binding description, the effective lattice spacing is the typical distance between two superconductors $W$.
Using a continuum description of the tight-binding model along the circumference, we get $\Delta_{\rm SO}\approx t_{\perp} \lambda^2$. 
Since $\lambda$ is the spin-dependent angle accumulated when hopping between nearest neighbors, and $\ell_{\rm SO}$ is the distance where a phase of $2\pi$ is accumulated, we have $W/\ell_{\rm SO}\approx\lambda$. Therefore, we obtain the condition 
\begin{equation}\label{eq:geometry_L}
    \frac{W}{\ell_{\rm SO}} \approx \frac{\Delta_{\rm SO}}{\Delta},
\end{equation}
which expresses the ideal geometry as a function of the continuum parameters only.

\begin{figure*}
    \centering\includegraphics[width=0.5\textwidth]{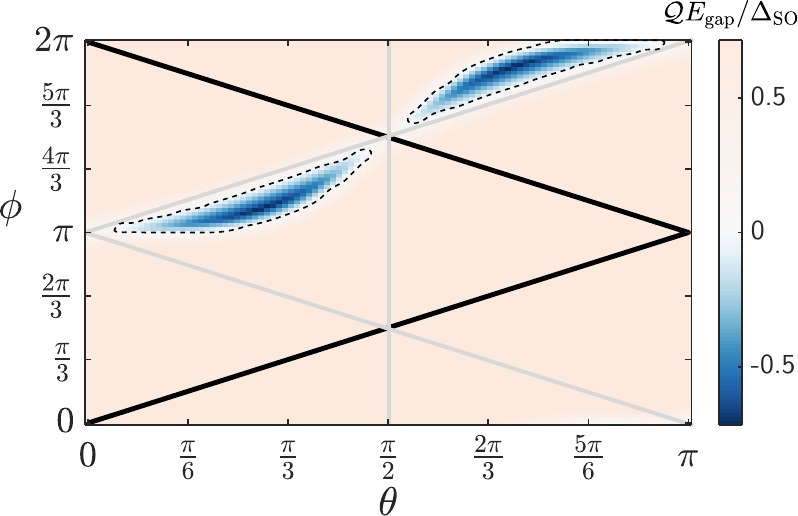}
    \caption{Topological phase diagram of the quantum-well model, same as in Fig.~\figref{fig:quantum_well}{b}, with $\Delta=0.5\,{\rm meV}$ and $W_{\rm SC}=100\,{\rm nm}$, $W_{\rm N}=100\,{\rm nm}$, $\mu=109.5\,{\rm meV}$, $t_{\perp}=0.28\,{\rm meV}$.}
    \label{sfig:Delta_0.5}
\end{figure*}

Let us exemplify the practical use of this relation. In the main text, we assumed a superconducting gap of $\Delta=1\,{\rm meV}$, which is appropriate for e.g. Nb and Pb~\cite{matthias_superconductivity_1963}. Let us now take $\Delta=0.5\,{\rm meV}$, which is appropriate for e.g. Sn and V~\cite{matthias_superconductivity_1963}. Using the relation Eq.~\eqref{eq:geometry_L} above, we simulate a larger system compared to that of Fig.~\figref{fig:quantum_well}{b}, with $W_{\rm SC}=100\,{\rm nm}$, $W_{\rm N}=100\,{\rm nm}$.
Fig.~\ref{sfig:Delta_0.5} shows the resulting topological phase diagram, which indeed exhibits topological regions with a topological gap comparable to $\Delta_{\rm SO}$, but the topological region is smaller, indicating that further optimization might be necessary.

\section{Bounds on the topological gap}
Here we discuss the bounds limiting the topological gap, in order to justify the choice of comparing it to $\Delta_{\rm SO}$, which we made in Fig.~\figref{fig:quantum_well}{b} of the main text.

To set the stage, we study the topological nanowire model~\cite{lutchyn_majorana_2010, oreg_helical_2010}
\begin{equation}\label{eq:nanowire}
    H=\left(\frac{k^{2}}{2m^{*}}+uk\sigma_{z}-\mu\right)\tau_{z}-B\sigma_{x}+\Delta\tau_{x},
\end{equation}
where $B$ is the applied Zeeman field and $u$ is the SOC parameter.
For simplicity, we focus on $\mu=0$ where the condition for a topological phase is $B>\Delta$.
The two relevant energy scales are $\Delta$ and $\Delta_{\text{SO}}=mu^{2}/2$, and the question is whether or not they both set a bound on the energy gap in the topological phase.

At finite $B$ and $\Delta$ there are two minima of the gap in the spectrum as a function of the momentum $k$, one at $k=0$ and the other near the Fermi momentum. The topological gap of the system is determined by the smallest of the two, when $B>\Delta$. 
It is maximized at $B=B^{*}>\Delta$, for which the gap at $k=0$ is equal to the gap near the Fermi momentum. 
A closed-form expression for $B^{*}$ is hard to obtain, but it is straightforward to find it numerically given the values of the other parameters. 

\begin{figure*}
    \centering
    \includegraphics[width=\textwidth]{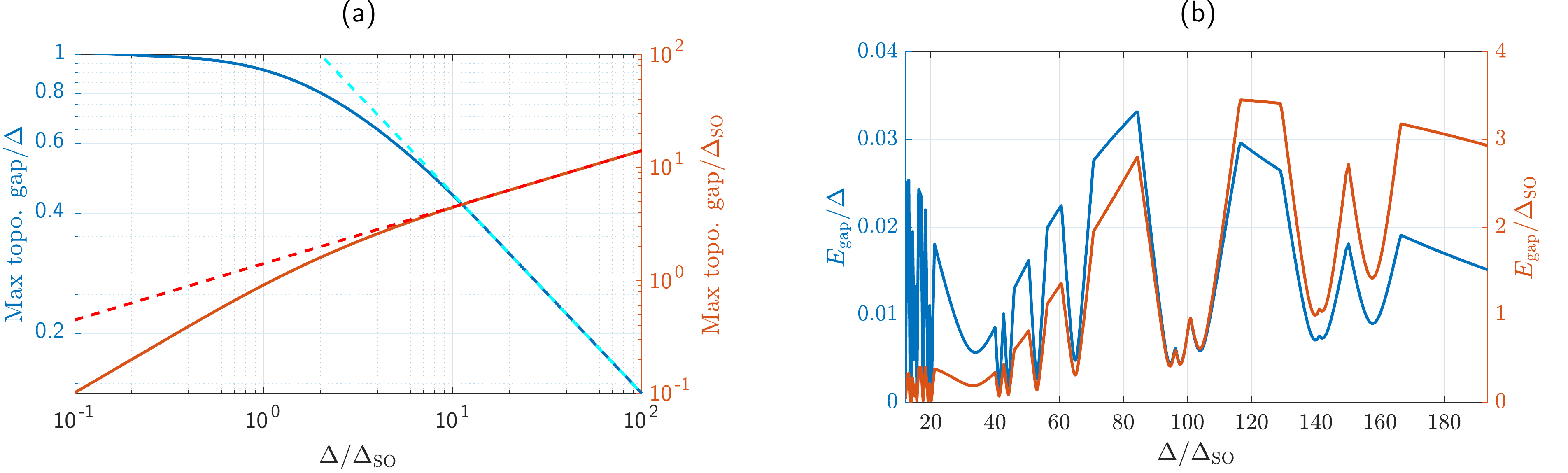}
    \caption{(a)~Maximal topological gap for the nanowire model Eq.~\eqref{eq:nanowire} (see also Refs.~\cite{lutchyn_majorana_2010,oreg_helical_2010}), as a function of $\Delta/\Delta_{\rm SO}$. The topological gap is normalized by $\Delta$ (blue) and by $\Delta_{\rm SO}$ (orange), and plotted in a log-log scale. The dashed lines are the asymptotic forms at $\Delta\gg\Delta_{\rm SO}$, which is $\sqrt{2\Delta_{\rm SO}\Delta}$, i.e., the topological gap may be parametrically larger than $\Delta_{\rm SO}$.
    (b)~Same for the quantum-well model Eq.~\eqref{eq:quantum_well}, using the same parameters of Fig.~\figref{fig:quantum_well}{b}. The maximal topological gap is of order $\Delta_{\rm SO}$, and since $\Delta\gg\Delta_{\rm SO}$ for the parameters we used, it is much smaller than $\Delta$.}
    \label{sfig:gap_bounds}
\end{figure*}

Fig.~\figref{sfig:gap_bounds}{a} shows the maximal topological gap as a function of $\Delta/\Delta_{\rm SO}$, normalized by $\Delta$ and by $\Delta_{\rm SO}$.
For InAs/InSb nanowires proximitized by Al, $\Delta$ and $\Delta_{\rm SO}$ are of the same order of magnitude.
However, for a InAs/InSb 2DEG such as the one we studied, $\Delta\gg\Delta_{\rm SO}$ and therefore we analyze the asymptotic behavior of the maximal topological gap in this limit---see the dashed lines in Fig.~\figref{sfig:gap_bounds}{a}.
By fitting the asymptotes, which can be obtained numerically or analytically, we find that the maximal topological gap in this limit is $\sim\sqrt{2\Delta_{\text{SO}}\Delta}$. 
Therefore, the gap can be parametrically larger than $\Delta_{\rm SO}$.
However, it is evident and also seen in Fig.~\figref{sfig:gap_bounds}{a} that the gap cannot exceed $\Delta$.

The situation is qualitatively different for the quantum-well model studied here, see Eq.~\eqref{eq:quantum_well} and Fig.~\ref{fig:quantum_well}. 
We demonstrate this by using the same parameters as in Fig.~\figref{fig:quantum_well}{b}, with the phases optimally chosen, and vary the ratio $\Delta/\Delta_{\rm SO}$.
The results are shown in Fig.~\figref{sfig:gap_bounds}{b}. 
It is clear from this figure that for our system, the maximal gap in the topological region is of order $\Delta_{\rm SO}$ (at the optimal configuration), but it is much smaller than $\Delta$.

\section{Stability to perturbations}

In this section, we analyze the stability of the topological phase in the quantum-well model to perturbations in the model's parameters. We demonstrate the robustness of the topological gap to various realistic imperfections, which makes our proposal favorable for experiments.

\begin{figure*}
    \centering
    \includegraphics[width=\textwidth]{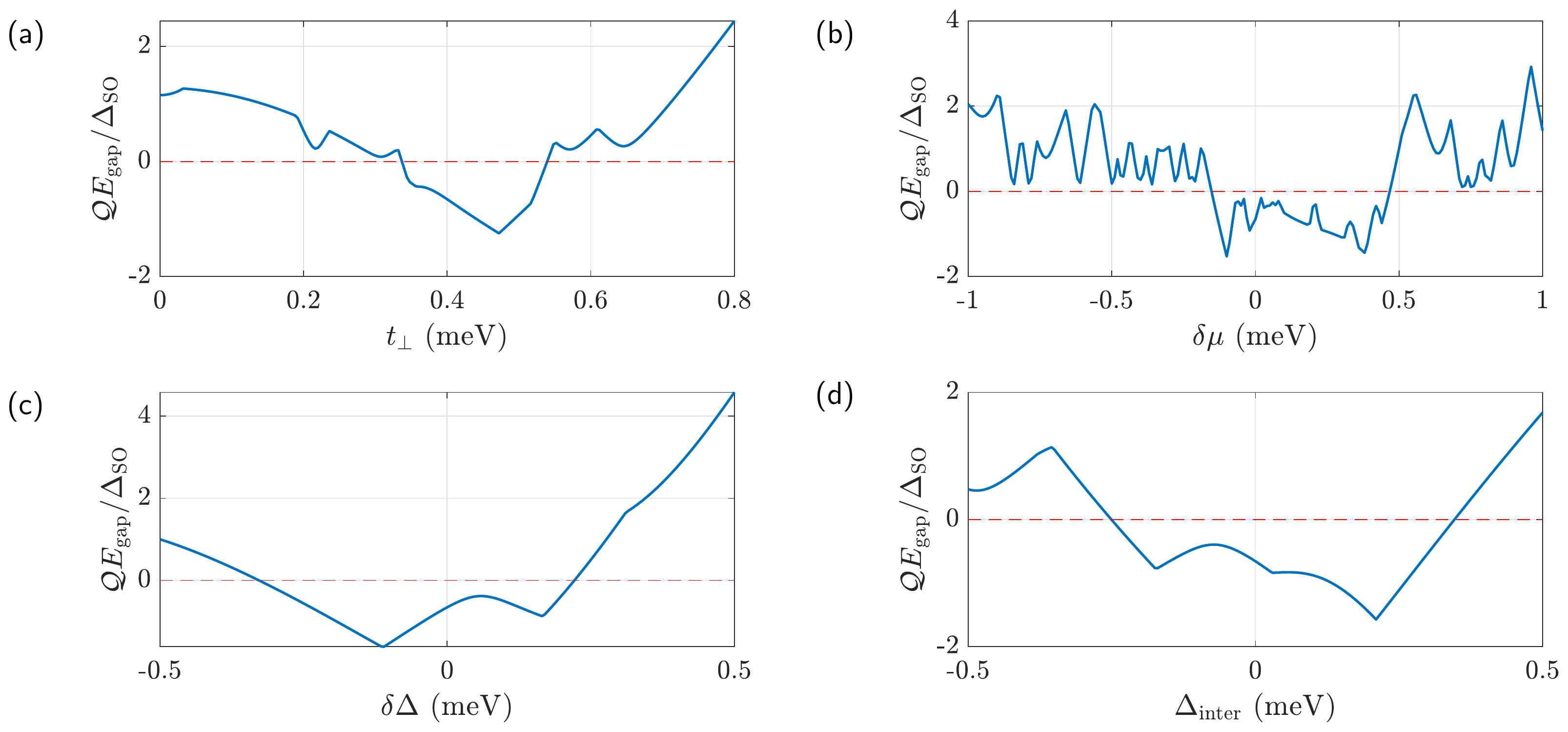}
    \caption{Stability of the topological phase in the quantum-well model to perturbations in (a)~the inter-layer hopping $t_{\perp}$, (b)~a difference $\delta\mu$ in the chemical potential between the two layers, (c)~a difference $\delta\Delta$ in the pair potential between the two layers, and (d)~inter-layer pair potential $\Delta_{\rm inter}$. Plotted is the topological invariant $\mathcal{Q}$ multiplied by the energy gap in units of the SOC energy. The dashed red line marks the topological phase boundaries.}
    \label{sfig:stability}
\end{figure*}

The parameters used in Fig.~\ref{sfig:stability} are the same as those of Fig.~\figref{fig:quantum_well}{b} of the main text, with $\theta=0.55\pi$, $\phi=0.88\pi$, a representative point inside the topological region. On top of these, we add perturbations as listed below. We plot the topological invariant $\mathcal{Q}$ multiplied by the energy gap, \emph{without} even changing the SC phases at all (which is probably the simplest experimental knob).

In Fig.~\figref{sfig:stability}{a}, the perturbation is a variation in the inter-layer hopping amplitude $t_{\perp}$.
In Fig.~\figref{sfig:stability}{b}, the chemical potential in the two layers is different: $\mu_{\rm top}=\mu + \delta\mu$, $\mu_{\rm bottom}=\mu - \delta\mu$.
In Fig.~\figref{sfig:stability}{c}, the pair potential in the two layers is different: in the main text we took $\Delta_{\rm top}=\Delta$, $\Delta_{\rm bottom}=0$, and now we take $\Delta_{\rm top}=\Delta+\delta\Delta$, $\Delta_{\rm bottom}=-\delta\Delta$.
Finally, in Fig.~\figref{sfig:stability}{d} we add an inter-layer pair potential $\Delta_{\rm inter}\tau_x \rho_x$.

Under all these perturbations, the topological phase is robust in an appreciable range of parameters. The important implication of this finding is that no fine-tuning is required to drive the system into the topological phase. We stress again that these reassuring results are obtained without further tuning of the SC phases, which will likely increase the stability even more.

\end{document}